\newcommand{\g}{$\gamma$}
\newcommand{\cd}{C$_6$D$_6$}
\newcommand{\cycling}{{\sc CYCLING}}
\newcommand{\near}{NEAR}
\documentclass[galaxies,article,accept,pdftex,moreauthors]{Definitions/mdpi} 

\firstpage{1} 
\pubvolume{1}
\issuenum{1}
\articlenumber{0}
\pubyear{2026}
\copyrightyear{2026}
\datereceived{ } 
\daterevised{ } 
\dateaccepted{ } 
\datepublished{ } 

\usepackage{svg}
\usepackage{graphicx}
\DeclareGraphicsExtensions{.pdf,.eps,.png,.jpg}
\usepackage{siunitx}

\Title{Direct Experiments of Neutron Capture on Stable and Unstable Isotopes for Stellar Nucleosynthesis Studies}

\Author{%
Jorge~Lerendegui-Marco$^{1,*}$\orcidA{}, %
Javier~Balibrea-Correa$^{1}$, %
Victor~Babiano-Suarez$^{1}$, %
C\'{e}sar~Domingo-Pardo$^{1}$, %
Gabriel~de la Fuente-Rosales$^{1}$, %
Bernardo~Gameiro$^{1}$, %
Ion~Ladarescu$^{1}$, %
Ariel~Tarife\~{n}o-Saldivia$^{1}$, %
Pablo~Torres-S\'{a}nchez$^{1}$, %
Oliver~Aberle$^{2}$
% ...
}

\AuthorNames{%
Jorge~Lerendegui-Marco,%
Victor~Babiano-Suarez,%
Javier~Balibrea-Correa,%
C\'{e}sar~Domingo-Pardo,%
Oliver~Aberle,%
Victor~Alcayne,%
Simone~Amaducci,%
Michael~Bacak,%
Jes\'{u}s~Bartolom\'{e},%
Aparna~Basavaraja-Allannavar,%
Ana-Paula~Bernardes,%
Eric~Berthoumieux,%
Roland~Beyer,%
Matthew~Birch,%
Selin~Birincioglu,%
Marian~Boromiza,%
Damir~Bosnar,%
Benedetta~Brusasco,%
Manuel~Caama\~{n}o,%
Aline~Cahuzac,%
Francisco~Calvi\~{n}o,%
Marco~Calviani,%
Daniel~Cano-Ott,%
Adri\`{a}~Casanovas,%
Donato~Castelluccio,%
Francesco~Cerutti,%
Gabriele~Cescutti,%
Enrico~Chiaveri,%
Gerardo~Claps,%
Paolo~Colombetti,%
Nicola~Colonna,%
Patrizio~Console Camprini,%
Guillem~Cort\'{e}s,%
Miguel~Cort\'{e}s-Giraldo,%
Luigi~Cosentino,%
Sergio~Cristallo,%
Angelica~D'Ottavi,%
Gabriel~de la Fuente Rosales,%
Maria~Diakaki,%
Mario~Di Castro,%
Augusto~Di Chicco,%
Mirco~Dietz,%
Emmeric~Dupont,%
Ignacio~Dur\'{a}n,%
Zinovia~Eleme,%
Sylvain~Fargier,%
Martin~Farkas,%
Beatriz~Fern\'{a}ndez-Dom\'{\i}nguez,%
Paolo~Finocchiaro,%
Will~Flanagan,%
Varvara~Foteinou,%
Valter~Furman,%
Bernardo~Gameiro,%
Aman~Gandhi,%
Francisco~Garc\'{\i}a-Infantes,%
Aleksandra~Gawlik-Rami\k{e}ga ,%
Gianpiero~Gervino,%
Simone~Gilardoni,%
Enrique~Gonz\'{a}lez-Romero,%
Styliani~Goula,%
Erich~Griesmayer,%
Carlos~Guerrero,%
Frank~Gunsing,%
Carlo~Gustavino,%
Jan~Heyse,%
William~Hillman,%
Elizabeth~Jacoby,%
David~Jenkins,%
Erwin~Jericha,%
Arnd~Junghans,%
Ulli~K\"{o}ster,%
Yacine~Kadi,%
Nasser~Kalantar-Nayestanaki,%
Kalliopi~Kaperoni,%
Myroslav~Kavatsyuk,%
Michael~Kokkoris,%
Sotirios~Kopanos,%
Yury~Kopatch,%
Milan~Krti\v{c}ka,%
Nikolaos~Kyritsis,%
Claudia~Lederer-Woods,%
Giuseppe~Lorusso,%
Alice~Manna,%
Trinitario~Mart\'{\i}nez,%
Marco~Mart\'{\i}nez-Ca\~{n}ada,%
Alessandro~Masi,%
Cristian~Massimi,%
Pierfrancesco~Mastinu,%
Mario~Mastromarco,%
Emilio-Andrea~Maugeri,%
Annamaria~Mazzone,%
Emilio~Mendoza,%
Alberto~Mengoni,%
Veatriki~Michalopoulou,%
Paolo~Milazzo,%
Jacob~Moldenhauer,%
Riccardo~Mucciola,%
Elizabeth~Musacchio Gonz\'{a}lez,%
Agatino~Musumarra,%
Alexandru~Negret,%
Emmanuel~Odusina,%
Dimitrios~Papanikolaou,%
Carlos~Paradela,%
Albert~Parmenter,%
Nikolas~Patronis,%
Jos\'{e}Antonio~Pav\'{o}n,%
Maria~Pellegriti,%
Pablo~P\'{e}rez-Maroto,%
Alberto~P\'{e}rez de Rada Fiol,%
Giulio~Perfetto,%
Jaros{\l}aw~Perkowski,%
Cristina~Petrone,%
Nicholas~Pieretti,%
Luciano~Piersanti,%
Elisa~Pirovano,%
Ignacio~Porras,%
Javier~Praena,%
Jos\'{e}-Manuel~Quesada,%
Ren\'{e}~Reifarth,%
Alejandro~Reina,%
Dimitri~Rochman,%
Yuriy~Romanets,%
Annie~Rooney,%
Gerard~Rovira,%
Carlo~Rubbia,%
Adri\'{a}n~S\'{a}nchez-Caballero,%
Nicol\'{a}s~S\'{a}nchez-V\'{a}zquez,%
Rudra~Sahoo,%
~Salma,%
Daniele~Scarpa,%
Gavin~Smith,%
Nikolay~Sosnin,%
Michele~Spelta,%
Krzysztof~Stasiak,%
Giuseppe~Tagliente,%
Antonella~Tamburrino,%
Ariel~Tarife\~{n}o-Saldivia,%
Diego~Tarr\'{\i}o,%
Pablo~Torres-S\'{a}nchez,%
Giorgios~Tsiledakis,%
Stanislav~Valenta,%
Pedro~Vaz,%
Gianfranco~Vecchio,%
Diego~Vescovi,%
Vasilis~Vlachoudis,%
Rosa~Vlastou,%
Anton~Wallner,%
Christina~Weiss,%
Tobias~Wright,%
Renjie~Wu,%
Roberto~Zarrella,%
Petar~\v{Z}ugec,%
}

\address{%
$^{1}$ \quad Instituto de F\'{\i}sica Corpuscular, CSIC - Universidad de Valencia, Spain \\
$^{2}$ \quad European Organization for Nuclear Research (CERN), Switzerland \\
$^{3}$ \quad Centro de Investigaciones Energ\'{e}ticas Medioambientales y Tecnol\'{o}gicas (CIEMAT), Spain \\
$^{4}$ \quad INFN Laboratori Nazionali del Sud, Catania, Italy \\
$^{5}$ \quad TU Wien, Atominstitut, Stadionallee 2, 1020 Wien, Austria \\
$^{6}$ \quad Universidad de Sevilla, Spain \\
$^{7}$ \quad Universitat Polit\`{e}cnica de Catalunya, Spain \\
$^{8}$ \quad CEA Irfu, Universit\'{e} Paris-Saclay, F-91191 Gif-sur-Yvette, France \\
$^{9}$ \quad Helmholtz-Zentrum Dresden-Rossendorf, Germany \\
$^{10}$ \quad University of Manchester, United Kingdom \\
$^{11}$ \quad School of Physics and Astronomy, University of Edinburgh, United Kingdom \\
$^{12}$ \quad Horia Hulubei National Institute of Physics and Nuclear Engineering, Romania \\
$^{13}$ \quad Department of Physics, Faculty of Science, University of Zagreb, Zagreb, Croatia \\
$^{14}$ \quad University of Santiago de Compostela, Spain \\
$^{15}$ \quad Agenzia nazionale per le nuove tecnologie, l'energia e lo sviluppo economico sostenibile (ENEA), Italy \\
$^{16}$ \quad Istituto Nazionale di Fisica Nucleare, Sezione di Bologna, Italy \\
$^{17}$ \quad Istituto Nazionale di Fisica Nucleare, Sezione di Trieste, Italy \\
$^{18}$ \quad Department of Physics, University of Trieste, Italy \\
$^{19}$ \quad INFN Laboratori Nazionali di Frascati, Italy \\
$^{20}$ \quad Istituto Nazionale di Fisica Nucleare, Sezione di Torino, Italy \\
$^{21}$ \quad Department of Physics, University of Torino, Italy \\
$^{22}$ \quad Istituto Nazionale di Fisica Nucleare, Sezione di Bari, Italy \\
$^{23}$ \quad Istituto Nazionale di Fisica Nucleare, Sezione di Perugia, Italy \\
$^{24}$ \quad Istituto Nazionale di Astrofisica - Osservatorio Astronomico d'Abruzzo, Italy \\
$^{25}$ \quad National Technical University of Athens, Greece \\
$^{26}$ \quad Physikalisch-Technische Bundesanstalt (PTB), Bundesallee 100, 38116 Braunschweig, Germany \\
$^{27}$ \quad University of Ioannina, Greece \\
$^{28}$ \quad University of Dallas, USA \\
$^{29}$ \quad Affiliated with an institute covered by a cooperation agreement with CERN \\
$^{30}$ \quad University of Lodz, Poland \\
$^{31}$ \quad Istituto Nazionale di Fisica Nucleare, Sezione di Roma1, Roma, Italy \\
$^{32}$ \quad European Commission, Joint Research Centre (JRC), Geel, Belgium \\
$^{33}$ \quad University of York, United Kingdom \\
$^{34}$ \quad Insitut Laue-Langevin, 71 Avenue des Martyrs, 38042, Grenoble, France \\
$^{35}$ \quad University of Groningen, NL-9747 AA Groningen, the Netherlands \\
$^{36}$ \quad Charles University, Prague, Czech Republic \\
$^{37}$ \quad National Physical Laboratory, Hampton Road, Teddington, UK \\
$^{38}$ \quad University of Granada, Spain \\
$^{39}$ \quad Dipartimento di Fisica e Astronomia, Universit\`{a} di Bologna, Italy \\
$^{40}$ \quad INFN Laboratori Nazionali di Legnaro, Italy \\
$^{41}$ \quad Dipartimento Interateneo di Fisica, Universit\`{a} degli Studi di Bari, Italy \\
$^{42}$ \quad Paul Scherrer Institut (PSI), Villigen, Switzerland \\
$^{43}$ \quad Consiglio Nazionale delle Ricerche, Bari, Italy \\
$^{44}$ \quad Istituto Nazionale di Fisica Nucleare, Sezione di Catania, Italy \\
$^{45}$ \quad Department of Physics and Astronomy, University of Catania, Italy \\
$^{46}$ \quad Goethe University Frankfurt, Germany \\
$^{47}$ \quad Instituto Superior T\'{e}cnico, Lisbon, Portugal \\
$^{48}$ \quad Japan Atomic Energy Agency (JAEA), Tokai-Mura, Japan \\
$^{49}$ \quad Department of Physics and Astronomy, Uppsala University, Box 516, 75120 Uppsala, Sweden \\
}

\corres{Correspondence: jorge.lerendegui@ific.uv.es}

\abstract{Neutron-capture reactions provide essential nuclear-physics input for modeling the synthesis of heavy elements in stars. The growing precision of stellar spectroscopy and isotopic measurements in presolar SiC grains now demands cross sections with improved accuracy over the full energy range, and access to unstable nuclei relevant to slow (s-) process branchings and the intermediate (i-) process. This article reviews recent progress in direct neutron-capture measurements, focusing on time-of-flight (TOF) experiments at CERN n\_TOF and complementary activation techniques. Substantial advances have been achieved for stable s-only and bottleneck isotopes, significantly improving constraints on s-process models. In parallel, the combination of high instantaneous neutron fluxes and advanced detector systems has facilitated first-time measurements of several radioactive branching-point nuclei. Feasibility studies, however, reveal current limitations related to sample availability, background conditions, and restricted energy coverage. In this context, the complementarity between TOF and activation emerges as a central strategy. Future developments, including high-flux facilities and novel inverse-kinematics experiments in ion storage rings, are expected to extend the boundaries of neutron-capture measurements, overcoming current limitations and helping to unlock new frontiers in our understanding of stellar nucleosynthesis.
}


\begin{document}

\section{Introduction}\label{Sec:Intro}

One of the most active research areas in nuclear astrophysics is the study of the nucleosynthesis of heavy elements (A $\gtrsim$ 60) in the universe. The production of these elements is primarily governed by neutron-capture processes, which play a fundamental role in shaping the observed abundance distribution of nuclei heavier than iron. The involved mechanisms, first presented in detail in the works of Burbidge~\cite{Burbidge:57} and Cameron~\cite{Cameron:57} almost 70 years ago, are characterized by different neutron densities of the astrophysical environment which lead to different relative timescales of neutron capture and $\beta$-decay.

The rapid neutron-capture process (r-process) occurs in explosive astrophysical environments, characterized by extremely high neutron densities (N$_n$ > 10$^{20}$ cm$^{-3}$), such as supernovae or neutron-star mergers. Under these conditions, successive neutron captures proceed much faster than $\beta$-decay, producing highly neutron-rich nuclei located far from the valley of stability. In contrast, the slow neutron-capture process (s-process) operates in stellar environments with relatively low neutron densities (N$_n$ $\approx$ 10$^{6}$-- 10$^{12}$cm$^{-3}$), such as asymptotic giant branch (AGB) stars during the late stages of stellar evolution~\cite{Kappeler11}. In this case, neutron capture occurs on timescales longer than or comparable to $\beta$-decay, and nucleosynthesis proceeds along the valley of stability through a sequence of neutron captures and $\beta$-decays, gradually producing heavier nuclei up to bismuth. This process is, together with the r-process, responsible for the production of the majority ($\approx$ 99\%) of the observed solar abundances of elements heavier than iron (see e.g. Fig. 1 of Ref~\cite{Dillmann:23}). More recently, an intermediate neutron-capture process (i-process), originally proposed by Cowan and Rose in 1977~\cite{Cowan77}, has been identified. This process occurs under neutron densities intermediate between those of the s- and r-processes (N$_n$ $\approx$ 10$^{13}$- 10$^{16}$cm$^{-3}$) and involves nuclei located only a few mass units away from stability. The i-process has been shown to play an important role in explaining the abundance patterns observed in certain low-metallicity AGB stars and carbon-enhanced metal-poor stars~\cite{Hampel16}. 

From the experimental nuclear physics perspective, the key ingredient to improve the nucleosynthesis models is the accurate determination of neutron-capture cross sections at all relevant energies. This article reviews recent developments in direct neutron-capture measurements relevant to s-process nucleosynthesis in AGB and massive stars~\cite{Kappeler11,Pignatari:10}. As described in Sec.~\ref{Sec:Methods}, two different methodologies for neutron-capture cross-section measurements have been extensively applied so far in many laboratories worldwide, neutron time-of-flight (TOF) and neutron activation, thereby covering more than 300 s-process involved nuclei, most of them stable~\cite{Dillmann:23,Domingo:25}. Despite the extensive experimental activities, there are still significant needs for new measurements, as discussed in Sec.~\ref{Sec:Needs}. Sec.~\ref{Sec:Highlights_nTOF} discusses the recent efforts at CERN n\_TOF to expand the experimental knowledge on direct neutron capture cross section. The main limitations of current state-of-the-art TOF experiments and the recent advances on both the TOF and activation techniques~\cite{Kappeler11} at CERN n\_TOF are then discussed in Sec.~\ref{Sec:Advances_Limits}. Lastly, Sec.~\ref{Sec:Future} presents future ideas to complement the existing methodologies with new measuring stations and techniques.


\section{Experimental methods for direct neutron capture measurements}\label{Sec:Methods}

\begin{figure}[!h]
 \centering
  \includegraphics[width=0.6\columnwidth]{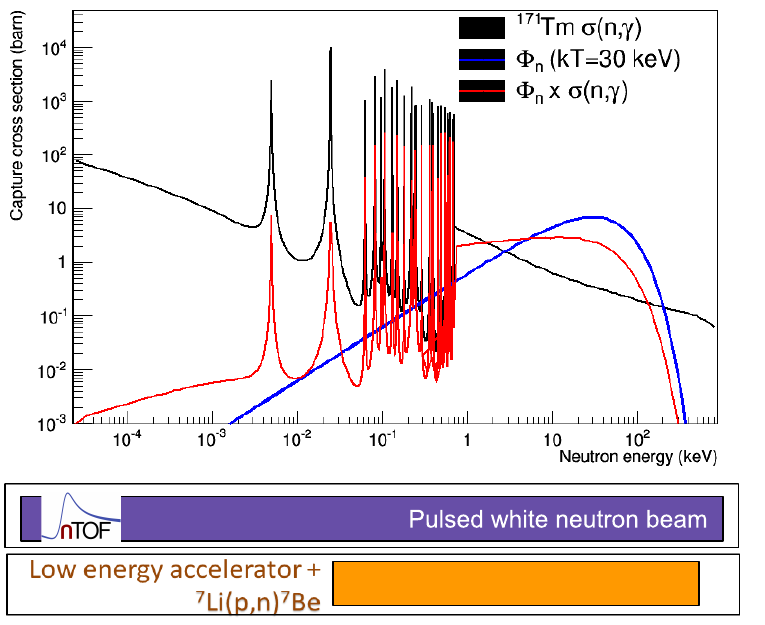} 
 \caption{Top: Maxwell-Boltzmann spectrum of neutrons at the reference stellar temperature $kT$~=~30~keV (blue curve), example of an (n,$\gamma$) cross section as available in JEFF-3.3 (black curve) and the convolution of the two (red curve), illustrating the contribution to the MACS. The units in the vertical axis are relevant only for the cross section. Bottom: Energy range covered by the two mostly employed beams and techniques for stellar (n,$\gamma$) measurements (see text for details). }
 \label{fig:XS_MACS_intro}
\end{figure}

In an astrophysical environment where nucleosynthesis occurs, the stellar plasma is characterized by a temperature T at which particles are in thermodynamic equilibrium and their energies follow a Maxwell-Boltzmann distribution. Consequently, the quantity of astrophysical relevance is the neutron-induced cross section averaged over this stellar neutron-energy distribution, commonly referred to as the Maxwellian-averaged cross section (MACS)~\cite{Massimi:22}. The latter has been compiled for s-process temperatures, $kT$~=~5--100~keV, in the KADoNiS~v0.3 database~\cite{KADONIS}. Fig.~\ref{fig:XS_MACS_intro} shows the example of the $^{171}$Tm(n,$\gamma$) cross section compared to the Maxwell-Boltzmann energy distribution at the reference temperature $kT$~=~30~keV (hereafter MACS$_{30}$) to illustrate the contribution of different energy ranges of the neutron capture cross section to the stellar capture rate.

Experimentally, neutron-capture cross sections relevant to stellar nucleosynthesis can be measured with direct methods using two complementary approaches: activation and time-of-flight (TOF) techniques. On the one hand, TOF facilities based today on spallation neutron sources produce white neutron spectra covering a broad neutron energy range. Combined with precise neutron-energy determination through the time-of-flight method, these facilities allow measurement of energy-dependent cross sections over a wide range of neutron energies, from which MACS values can be derived for astrophysical relevant temperatures. Neutron capture measurements for astrophysics are carried out in different white-neutron beam facilities across the world, such as GELINA~\cite{GELINA}, LANSCE~\cite{DANCE}, ANNRI~\cite{ANNRI}, Back-n~\cite{BACKN} or n\_TOF~\cite{Guerrero13,Pavon:25}. Sec.~\ref{Sec:Highlights_nTOF} will focus on the latter and review some of the recent highlights on neutron capture measurements of astrophysical relevance.

Neutron-capture TOF experiments are sometimes limited by the quality (isotopic enrichment) and quantity of sample material available for the experiment. This is particularly critical for the case of unstable isotopes, such as the s-process branching points~\cite{Guerrero:17}. Alternatively, in activation experiments, samples are irradiated with neutron spectra resembling stellar neutron distributions, allowing direct determination of MACS values at specific temperatures. As indicated in the bottom panel of Fig.~\ref{fig:XS_MACS_intro}, a quasi-stellar neutron spectrum at $kT$~=~25~keV can be obtained using a low-energy proton accelerator and the $^{7}$Li(p,n) reaction~\cite{Wallner:16}. An intense research program on direct (n,\g) activation measurements for astrophysics has been carried out in facilities like FZK~\cite{Reifarth:03}, LiLiT-SARAF~\cite{LILIT-SARAF}, HISPANoS-CNA~\cite{MillanCallado:24} or LeNoS-LNL~\cite{Mastinu:12}.

The activation technique in world-leading facilities in terms of neutron flux, FZK~\cite{Reifarth:03,Wallner:16} and LiLiT-SARAF~\cite{Guerrero:2020}, has shown an unsurpassed sensitivity for the measurement of sub-microgram samples (i.e. 10$^{14}$ atoms)~\cite{Reifarth:03} and can be the only direct method for cases where the high sample-decay induced background would represent an important limitation for the TOF measurement (see Sec.~\ref{Sec:Highlights_nTOF}). On the contrary, activation in quasi-stellar beams has been typically limited to a single accessible stellar temperature ($kT$~=~25~keV). Several facilities have studied methodologies to tune the outgoing neutron spectrum by modifying the proton energy and sample angle~\cite{Mastinu:12,Musachio:24,Praena:24}, using ring-shaped sample to access lower s-process temperatures~\cite{Heftrich:23} or combining multiple proton energies to access higher temperatures ($kT$~=~90keV) found in massive stars~\cite{PerezMaroto:24}.

With the two-fold aim of covering a wider range and enabling direct measurement of low mass samples, mainly of unstable species not accessible via TOF, a new generation of high-flux neutron-activation stations with tuneable energy range has been proposed at CERN~\cite{Patronis:23,BDF_Proposal}, as it is presented in Secs.~\ref{Sec:Advances_Limits} and \ref{Sec:Future}. 

\section{Data needs for stable and unstable isotopes}\label{Sec:Needs}

\begin{figure}[!h]
 \centering
  \includegraphics[width=0.9\columnwidth]{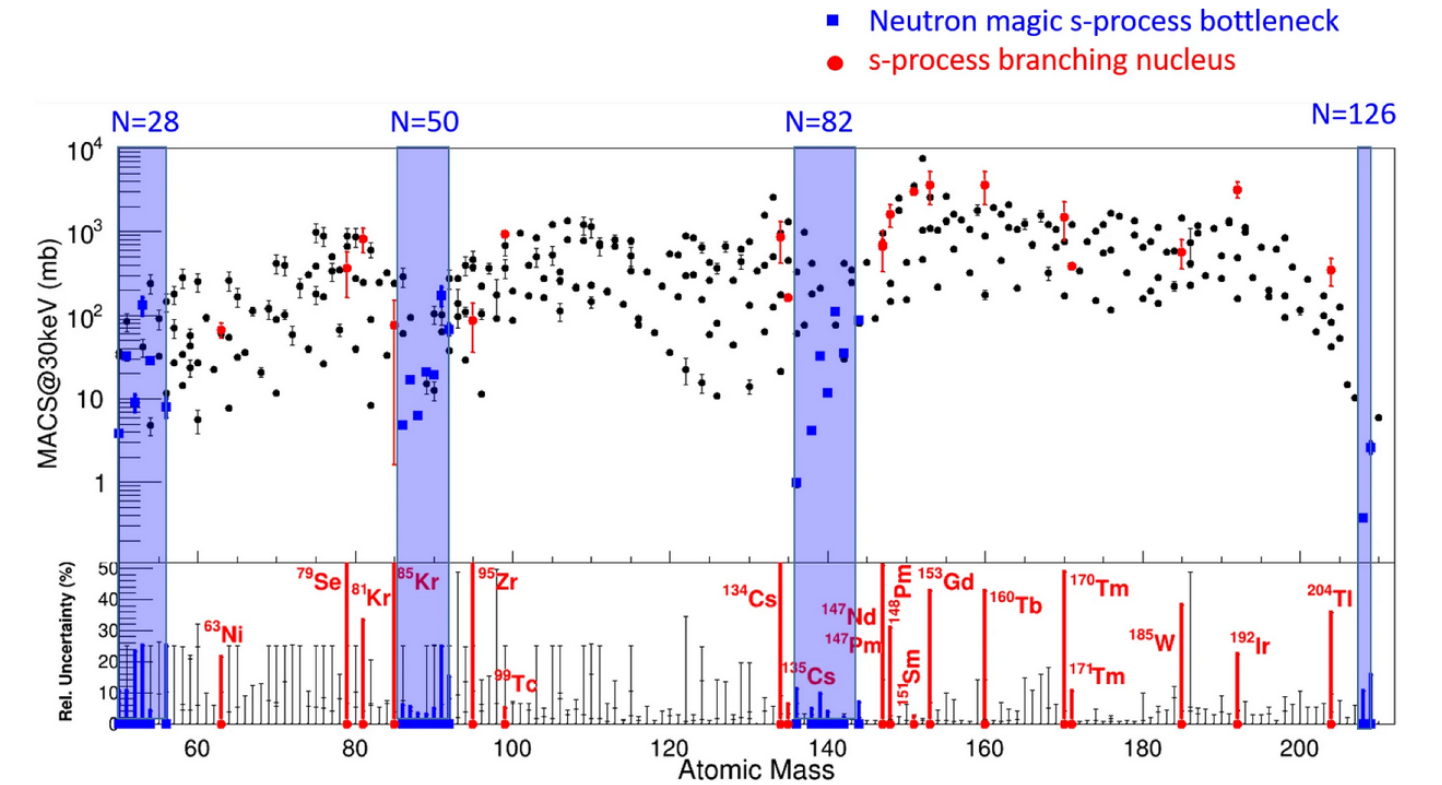}
 \caption{MACS$_30$ in KADoNis v0.3~\cite{KADONIS} (top) and their present accuracy (bottom). Nuclei acting as s-process bottlenecks and branching points are displayed in blue and red, respectively. Figure taken from Ref.~\cite{Domingo:22}.}
 \label{fig:MACS_summary}
\end{figure}

Despite the extensive experimental activities over the last decades, there are still acute needs for new neutron-capture cross-section measurements, and for a large number of the measured isotopes improvements are required both in the covered energy range (approximately from 1~eV up to 100~keV) and in the obtained accuracy. Related to the latter, many previous measurements present cross-section uncertainties that are significantly larger than the 5\% attainable in the measured abundance ratios from meteorite grains~\cite{Kappeler11}. The values of MACS$_{30}$ for all the nuclei involved in the s-process pathway and their current uncertainties are depicted in Fig.~\ref{fig:MACS_summary}. Among all the isotopes plotted in this figure, several groups of key nuclei can be distinguished which have a crucial impact on the nucleosynthesis process~\cite{Dillmann:23,Domingo:22,Domingo:25}. An accurate knowledge of their stellar capture rate is indispensable for a better understanding of the neutron economy in the s-process and the resulting calculated abundances. The present status and needs for these particularly relevant isotopes are summarized in Table~\ref{tab_dataneeds}.

The first family of key isotopes are the s-only isotopes, nuclei which are solely produced by the s-process since they are blocked from r-process contributions by other stable isobars. The s-only abundances serve as a benchmark for the nucleosynthesis and galactic-chemical evolution models. The MACS of the s-only nuclei are known with few exceptions within the required 5\% or even down to 2\%. A complete list of the 33 s-only isotopes and their current uncertainties can be found in Ref~\cite{Dillmann:23}.

Further, nuclei located in neutron shell-closures, characterized by small capture cross sections (see Fig.~\ref{fig:MACS_summary}), act as s-process bottlenecks. The s-process flow gives rise to the three characteristic abundance peaks at the neutron-shell closures. These elements --- Sr, Y, Zr (N = 50), Ba, La, Ce, Nd, Sm (N = 82) and Pb, Bi (N = 126)--- show-up prominently in spectroscopic observations of stellar atmospheres and thus, they represent a sensitive probe for stellar models. However, as a consequence of their small cross sections, experimental uncertainties are still large (even higher than 10\%) for some neutron-magic nuclei~\cite{Dillmann:23,Domingo:25}.

Particular attention has also been given to unstable nuclei acting as branching points in the s-process, highlighted in red in Fig.~\ref{fig:MACS_summary}. At these nuclei, featuring relatively long half-lives (years), neutron capture and $\beta$-decay compete, leading to a branching of the nucleosynthesis path~\cite{Kappeler11}. The resulting isotopic abundance patterns are highly sensitive to the physical conditions of the stellar environment, such as temperature and neutron density. Therefore, precise measurements of neutron-capture cross sections at branching points, combined with abundance observations and stellar models, provide valuable constraints on stellar evolution and nucleosynthesis conditions. 

Despite their relevance, only a small fraction of the key s-process branching nuclei listed in ~\cite{Kappeler11} have been measured with high accuracy to date~\cite{Guerrero:17,Domingo:22} and there is still a significant number of them that have not been accessed owing to several experimental limitations. The latter comprise the difficulty to produce samples of these isotopes in sizable amounts with a high enough enrichment. Furthermore, their activity implies a considerable radiation hazard and represents an intense source of background. As discussed in Sec.~\ref{Sec:Highlights_nTOF}, a combined effort has been carried out during the last decade between Institut Laue-Langevin (ILL)-Grenoble (France) and Paul Scherrer Institute (PSI)-Villigen (Switzerland) to produce high quality samples of these nuclei to measure the neutron capture cross section via TOF at CERN n\_TOF~\cite{Lerendegui:18b}. 

\begin{table}[H] 
\caption{Summary of data needs on direct (n,$\gamma$) measurements for key isotopes for stellar nucleosynthesis studies.\label{tab_dataneeds}}
\begin{tabularx}{\textwidth}{CCCCC}
\toprule
\textbf{Process}	& \textbf{Group of nuclei}	& \textbf{Challenge} & \textbf{Status} & \textbf{Main need}\\
\midrule
s-process	& s-only isotopes	& Required accuracy & Unc. $\geq$5\% (e.g. $^{80,82}$Kr, $^{198}$Hg) & Higher accuracy \\
s-process	& bottlenecks		& Small $\sigma$  & Unc. $\geq$10\% (e.g. $^{208}$Pb) & Higher sensitivity \\
s-process	& branching points	& Low mass, activity background & Few measured & Higher flux\\
\midrule
i-process	& See e.g. Refs.~\cite{Mckay:19,Denissenkov:21,Martinet:24}  & Short-lived unstable  & No data   & New concepts (See Sec.~\ref{Sec:Future})\\

\bottomrule
\end{tabularx}
\end{table}

Lastly, neutron-capture reactions on radioactive isotopes also play a crucial role in the i-process nucleosynthesis. The reaction flow of the i-process travels through radioactive nuclei 2--6 units away from stability on the neutron-rich side, where no experimental data from direct neutron capture cross sections are available to date. Recent Monte Carlo sensitivity studies~\cite{Mckay:19,Denissenkov:21,Martinet:24} have identified relevant neutron capture reactions which significantly impact i-process abundance predictions, such as $^{66}$Ni~(55~h), $^{72}$Zn~(47~h), $^{135}$I~(6.6~h), $^{153}$Sm~(46~h), $^{217}$Bi~(98.5~s), all featuring very short half-lives, which makes them inaccessible for (n,\g) measurements via direct methods. Other relevant cases related to i-process tracers such as $^{137}$Cs~(30~y), $^{144}$Ce~(285~d)~\cite{Choplin:21} may become feasible sooner. In this context, future high-flux neutron sources and innovative experimental methods, may make possible to address the direct neutron capture measurements of such reactions for the first time, as discussed in Sec.~\ref{Sec:Future}.


\section{Highlights of stellar neutron capture measurements at CERN n\_TOF} \label{Sec:Highlights_nTOF}
\subsection{Neutron capture measurements at the n\_TOF Facility}
The n\_TOF facility at CERN generates its neutron beams through spallation reactions of 20~GeV/c protons extracted in pulses from the CERN Proton Synchrotron and impinging onto a lead spallation target. The resulting high energy (MeV-GeV) spallation neutrons are partially moderated in a surrounding water layer to produce a white-spectrum neutron beam that expands in energy from thermal to a few GeV. The neutrons travel along two beam lines towards two experimental areas: EAR1 at 185~m (horizontal)~\cite{Guerrero13} and EAR2 at 19~m (vertical)~\cite{Weiss15,Pavon:25}. A new experimental area, so-called NEAR Station, has been installed during the latest upgrade next to the spallation target, as it is discussed in Sec.~\ref{Sec:Advances_Limits}. It features an extremely high neutron flux that opens the door to (n,\g) activation measurements on short-lived radioactive isotopes or on very small mass samples~\cite{Patronis:23}. 

In neutron TOF capture measurements at n\_TOF the sample of the isotope of interest is placed in the pulsed neutron beam and the prompt capture $\gamma$-rays originating from the sample are registered by means of radiation detectors. Two different detection systems have been used for (n,\g) measurements at n\_TOF so far: the 4$\pi$ Total Absorption Calorimeter (TAC)~\cite{Guerrero:09}--- a segmented detector array consisting of 40 BaF$_{2}$ crystals--- and low-efficiency \cd~liquid scintillators~\cite{Plag03} in conjunction with the pulse-height weighting technique (PHWT)~\cite{Tain02,Tain04}, which allows one to virtually mimic an ideal total energy detector (TED)~\cite{Moxon63}. The latter have been used for measurements of stellar MACS due to their reduced neutron sensitivity and fast response. Thus, all the measurements presented in this work rely on the TED principle. 

Since 2001, more than 60 neutron capture cross section measurements have been carried out at the first experimental area n\_TOF EAR1~\cite{Guerrero13}. Thanks to its long flight path of 185~m, EAR1 has been a reference facility for high resolution measurements on stable isotopes and samples with sufficient masses ($\geq$100~mg) both in the resonance region (RRR)~\cite{Lerendegui:18,BabianoND:22} and Unresolved Resonance Region (URR) up to 1~MeV~\cite{Aerts:06,Lederer:11,Mingrone:17,Lerendegui:26}, covering the full energy range of importance for astrophysics experiments. In 2014, the n\_TOF Collaboration built the vertical beam line n\_TOF-EAR2~\cite{Weiss15,Pavon:25}, featuring a flight path of only 20~m, which delivers a substantially higher neutron flux and is particularly advantageous for measurements on radioactive samples.

\subsection{Stable nuclei: s-process bottlenecks and s-only isotopes}

 A large fraction of the neutron-capture program at CERN n\_TOF over the past 25 years has focused on nuclear astrophysics~\cite{Massimi:22,Milazzo:25,Domingo:25}, addressing several of the key stable isotopes mentioned in Sec.~\ref{Sec:Needs}, namely s-process bottleneck and s-only isotopes, and contributing to solve discrepancies between SiC grain data and model predictions.

Among other measurements, the capture cross-section of the s-only $^{154}Gd$ provides a direct probe of stellar neutron exposures and the efficiency of neutron sources in AGB stars, particularly the structure and extent of the $^{13}$C pocket, which governs the main neutron production through the $^{13}$C($\alpha$,n)$^{16}$O reaction. Aiming at solving the observed underproduction of $^{154}$Gd relative to neighboring s-only isotopes such as $^{150}$Sm, a new measurement was carried out at n\_TOF EAR1~\cite{Mazzone:20} using low neutron-sensitivity \cd~detectors. The resulting MACS$_{30}$ = 880(50)~mb, significantly lower than previously recommended values, leads to an increase of about 10\% in the predicted stellar abundance of $^{154}$Gd and improves the agreement between theoretical nucleosynthesis models and observed isotopic ratio of  $^{154}$Gd/$^{150}$Sm in the solar system.

Several recent experiments also cope with s-process bottlenecks, such as $^{140}$Ce or $^{209}$Bi. $^{140}$Ce lies in the second s-process peak (N = 82). The relatively low neutron-capture cross section of isotopes in this region, gives rise to the observed abundance peak around A $\approx$ 140. Precise neutron-capture measurements on $^{140}$Ce were required to solve the existing discrepancy of 30\% found between the Ce abundance predicted with AGB models and the spectroscopic observations~\cite{Straniero:14}. A new measurement of this cross section was carried out at n\_TOF with a highly enriched 12.3~g (Ce-oxide) sample produced at PSI. The high-resolution measurement using the aforementioned \cd~setup with low neutron sensitivity covered a large neutron-energy range up to 65 keV. The relevant MACS for nucleosynthesis in AGB stars, at $kT$~=~8~keV, was found to be of 28.18(24) mb, 40\% higher than expected ~\cite{KADONIS}. These results yielded an even smaller prediction in the s-process abundance of Ce with respect to previous estimations. 

At the termination of the s-process path, $^{209}$Bi represents the heaviest stable isotope and plays a key role in determining the final abundance distribution of heavy elements produced by neutron capture. The measurement of its MACS, featuring a very low value of only 2.9(5) mb at $kT$~=~25 keV, together with the cross sections of the neighboring Pb isotopes (see e.g.~\cite{Domingo:06_b}), govern the s-process termination. In particular, rather accurate abundance constraints could be derived for the r-process contribution to $^{209}$Bi. With this motivation, a first neutron-capture TOF measurement on $^{209}$Bi was carried out at EAR1 in 2001~\cite{Domingo06}. However, it encountered many difficulties due to the in-beam \g-ray background and limited statistics. To overcome these limitations, a new measurement was successfully carried out in 2024 at n\_TOF EAR2~\cite{Bi209Proposal}. The high-flux at EAR2 allowed to improve the covered energy range and the statistical accuracy with the same sample used in EAR1. Beyond this, it made possible the measurement of a new thinner (1~mm) sample for a better assessment of multiple-scattering and neutron-sensitivity effects. Moreover, the total cross section has been measured by means of transmission at JRC-Geel~\cite{Bi209_Geel}, thereby allowing a fully consistent R-matrix analysis of the two data sets.

\begin{figure}[!h]
 \centering
  \includegraphics[width=1.0\columnwidth]{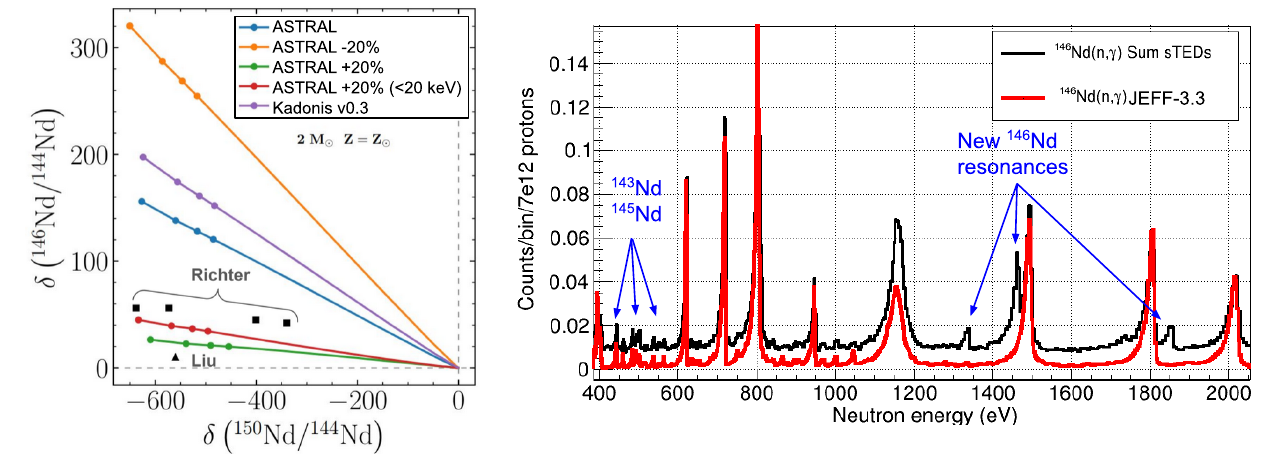}
 \caption{Left: $\delta$-values (per mil deviations from solar system ratios) of $^{146}$Nd/ $^{144}$Nd relative to $^{150}$Nd/ $^{144}$Nd from AGB star calculations (solid lines) compared with SiC grain data (black dots) from Richter~\cite{Richter:92} and Liu~\cite{LiuPhD}), showing the impact of varying the $^{146}$Nd(n,\g) rate of ASTRAL~\cite{ASTRAL} in $\pm$20\% in the full energy range and only below $kT$=20~keV. Right: Example of the counting rate measured at n\_TOF EAR2 in a small range of neutron energies compared to the expected results based on JEFF-3.3.}
 \label{fig:Nd146}
\end{figure}

Beyond s-only and bottleneck isotopes, an active program at n\_TOF is also addressing present discrepancies between isotopic abundances measured in presolar SiC grains and the predictions of state-of-the-art s-process nucleosynthesis models. One of the recent cases is $^{146}$Nd. Persistent discrepancies are observed between the predicted and measured $^{146}$Nd/ $^{144}$Nd ratio, which is systematically overestimated by current models, when compared to SiC data~\cite{Yin:06}. Sensitivity studies carried out with the FUNS code~\cite{Cristallo:15} indicate that an increase of about 15\% in the $^{146}$Nd neutron-capture cross section only at low stellar temperatures (below $kT$=20~keV), would significantly improve the agreement with the grain data (see Fig.~\ref{fig:Nd146}). The available experimental information is largely based on activation and time-of-flight data covering only the region above 3 keV~\cite{Whissak:98}, so-called unresolved resonance region (URR). In contrast, the capture cross section at lower energies in the so-called resolved resonance region (RRR), which strongly impacts the MACS at low stellar temperatures ($kT$~$\approx$~8~keV), has never been measured. To address this issue, a  high-resolution TOF experiment was performed at n\_TOF EAR2 using a 212 mg $^{146}$Nd oxide sample enriched to 97.4\%, aiming at measuring both the RRR up to 5 keV and the URR up to 100~keV~\cite{INTC_Nd146}. The TOF measurement at EAR2, for which preliminary results are shown in Fig.~\ref{fig:Nd146}, has been complemented with an activation experiment at HiSPANoS-CNA to determine the MACS at $kT$~=~25 keV. Last, this measurement has become the first case at n\_TOF exploiting the complementarity between a TOF measurement and activation measurements in the new high-flux n\_TOF-NEAR facility~\cite{INTC_Nd146_2}, thereby establishing a benchmark for future experiments made with this combination of both approaches.

\subsection{Unstable nuclei: s-process branching points}

Experimental efforts at CERN n\_TOF have also focused on the more challenging measurement of unstable nuclei which act as branchings of the s-process. The limited sample mass available and the high background induced by the sample activity represent the major challenges to experimentally access the (n,\g) cross sections of these isotopes. To cope with such demanding requisites, a big effort has been carried out to improve the neutron beam --- enhanced luminosity and improved resolution --- and to reduce the $\gamma$- and neutron-induced backgrounds~\cite{Lerendegui:23_ND,Domingo:22,Domingo:25}. In parallel, the collaboration has developed an active R\&D program on innovative detection systems~\cite{Mastinu:13,Domingo16,Babiano:21,Alcayne:24,Balibrea:25}, leading to a progressive increase in detection sensitivity for the radiative neutron-capture, that has allowed to address several first-ever TOF measurements on key s-process branchings. The reader is referred to Ref.~\cite{Domingo:25} for an extended discussion on the evolution of the difficulty level of the measurements on unstable nuclei. 

The early TOF measurements on unstable nuclei took place in n\_TOF EAR1 between 2000 and 2015. In particular, the capture cross sections of $^{151}$Sm~\cite{Abbondanno:04}, $^{93}$Zr~\cite{Tagliente13}, $^{63}$Ni~\cite{Lederer:13},$^{171}$Tm~\cite{Guerrero:2020}, $^{204}$Tl~\cite{Casanovas:24} were measured. Despite the major progress related to the neutron beams and improvement in the detection sensitivity along the years, these experiments in EAR1 showed some limitations. In particular, the $^{93}$Zr~\cite{Tagliente13} and $^{63}$Ni~\cite{Lederer:13} measurements suffered from a low signal-to-background ratio (SBR) due to the neutron-induced background, which limited the neutron energy range that could be measured to few keV. More recently, the first ever TOF measurements of $^{171}$Tm~\cite{Guerrero:2020} and $^{204}$Tl~\cite{Casanovas:24} were carried out with samples containing only few mg ($\approx$10$^{19}$ atoms) produced in collaboration between ILL and PSI. The astrophysical impact of these experiments was limited by the very intense background arising from the sample activity, which restricted the upper energy for the analysis of resonances to 4~keV for $^{204}$Tl and only 700~eV for $^{171}$Tm. The latter is one of the relatively few cases where the TOF measurement could be complemented with an activation experiment at LiLiT-SARAF which, in turn, became crucial for improving the experimental uncertainty of MACS$_{30}$ to about 10\%. In practice, the idea of complementing TOF and activation was among the motivations to build the n\_TOF-NEAR station, discussed in Sec.~\ref{Sec:Advances_Limits} and the project for the future n\_ACT (see Sec.~\ref{Sec:Future}).

  A major upgrade to address the challenging background arising from radioactive samples was the deployment of EAR2. Thanks to its $\approx$10-times shorter flightpath compared to EAR1, it features a 400-times higher instantaneous neutron flux (see Ref.~\cite{Lerendegui16}). This beamline is particularly well suited for neutron-capture measurements on highly radioactive and/or very small mass samples, such as s-process branching nuclei. Later, in 2021, the n\_TOF facility installed its third generation spallation target~\cite{Exposito:21} during the CERN Long Shutdown 2 (LS2) (2019--2021). The optimized design of this new spallation target brought a remarkable upgrade of the performance of EAR2, both on the energy resolution and the neutron flux~\cite{Lerendegui:23_ND}. Details on the upgraded facility performance can be found in Ref.~\cite{BacakND:22,PavonND:22}. 

\begin{figure}[!h]
 \centering
  \includegraphics[width=1.05\columnwidth]{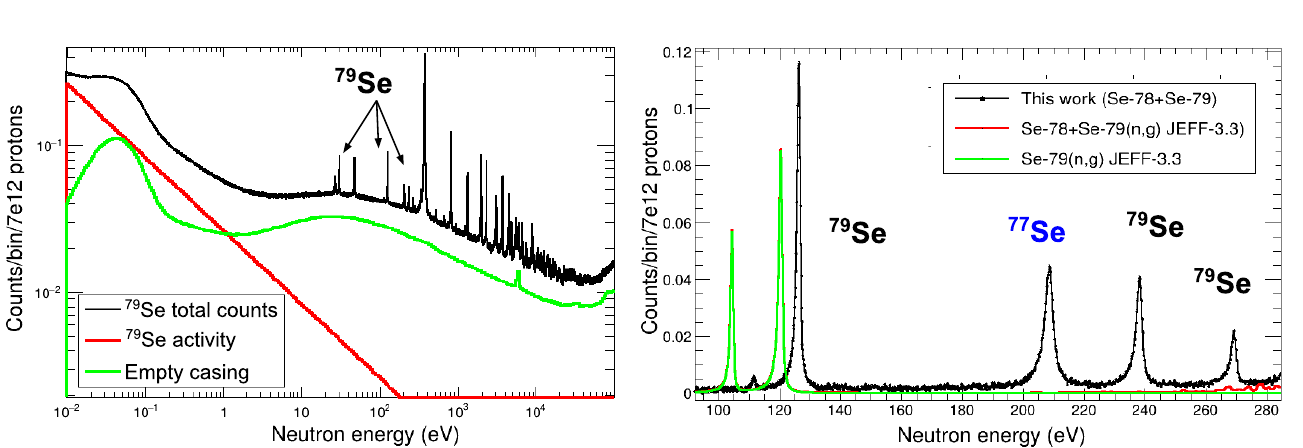}
 \caption{Left: Total counts and background components as a function of the neutron energy measured with the PbSe~($^{78}$Se+$^{79}$Se) sample in n\_TOF EAR2. The first resonances of $^{79}$Se are highlighted. Right: Background-subtracted $^{79}$Se(n,$\gamma$) yield showing examples of observed resonances compared to the expected results based on JEFF-3.3.}
 \label{fig:Se79}
\end{figure}

Improving the characteristics of the neutron beam is not sufficient to overcome all the existing challenges for (n,\g) measurements on several unstable branching points. To this end, together with the unprecedentedly high luminosity of EAR2 target, a new generation of \cd~detectors with higher segmentation, so-called sTED~\cite{Alcayne:24}, have been developed to cope with the high count rates associated to the flux of EAR2~\cite{Lerendegui:23_NPA}. The arrangement of the 9~sTED cells in a compact-ring configuration around the capture sample has enhanced the sensitivity~\cite{Lerendegui:23_NPA,Balibrea:23}. In addition, for those cases where the neutron-induced background dominates~\cite{Lederer:13,Tagliente13}, an innovative solution based on Total-Energy Detection (TED) with $\gamma$-ray imaging capability, so-called i-TED, was proposed~\cite{Domingo16}. i-TED exploits the Compton imaging technique with the aim of determining the direction of the incoming $\gamma$-rays. This allows the rejection of events not originating in the sample, thereby enhancing the signal-to-background ratio (SBR). i-TED has been developed~\cite{Babiano19,Babiano20,Balibrea:21} and  successfully validated at n\_TOF~\cite{Babiano:21,Lerendegui:22_NIC} in recent years.

The recent upgrades on detection systems and spallation target have allowed to reach an unprecedented sensitivity (see Fig. 2 of Ref.~\cite{Domingo:25}), as it is demonstrated by the first-ever capture measurements on the unstable $^{94}$Nb~(half-life, T$_{1/2}$~=~\SI{2.3e4}{y})~\cite{Balibrea:23,Balibrea:26} and $^{79}$Se~(T$_{1/2}$~=~ \SI{3.27e5}{y})~\cite{Lerendegui:23_NPA} isotopes. Focusing on the latter, the branching at $^{79}$Se is particularly well suited for determining the thermal conditions of the stellar environment thanks to the strong thermal dependence of its beta decay rate~\cite{Kappeler11}. For this experiment, 2.7~mg of $^{79}$Se were produced by means of neutron irradiation of an enriched $^{208}$Pb$^{78}$Se alloy-sample in the high-flux reactor at ILL~\cite{Chiera22,Lerendegui:23_NPA}. This measurement combined EAR1 using i-TED and EAR2 using the sTED array~\cite{Lerendegui:23_NPA}. The data analysis has shown the relevance of having a high instantaneous flux of EAR2 to cope with the background arising from unavoidable activation in the sample (5 MBq and 1.4 MBq for $^{75}$Se and $^{60}$Co, respectively), see Fig.~\ref{fig:Se79}. The quality of the TOF data has allowed analyzing for the first time 12--13 resonances up to $\approx$1.25~keV. So far, there are only theoretical predictions of the cross section, as those adopted in JEFF-3.3. However, this prediction can never be reliable as illustrated in Fig.~\ref{fig:Se79}. The final analysis will provide the first experimental value for the MACS, leading to a constraint of the thermal conditions of the weak s-process in massive stars.

\section{Current limits and recent advances} \label{Sec:Advances_Limits}
\subsection{Pushing the limits for TOF measurements in EAR2}

Neutron capture measurements on unstable isotopes have been already successfully measured at n\_TOF using samples with masses $\simeq$1-50~mg~\cite{Guerrero:2020,Casanovas:20,AlcayneND:22,Balibrea:23,Lerendegui:23_NPA}. Measurements of lower masses are still difficult, for instance a $^{147}$Pm(n,\g) measurement at EAR2 on a sample of only 85~$\mu$g, allowed to see only the 3 first resonances due to the limited SBR~\cite{Lerendegui:18b,Guerrero:19}.

 The recent improvements in the performance of the n\_TOF neutron beams and detection systems has led to an improved sensitivity in (n,\g) measurements. In this context, a systematic study of the current detection limit in the upgraded n\_TOF EAR2 has been carried out~\cite{Lerendegui:25b} for TOF measurements on key s-process branching isotopes listed in Refs.~\cite{Kappeler11,Guerrero:17}. Among the challenging key s-process points never measured before via TOF~\cite{Kappeler11,Guerrero:17}, we have selected $^{81}$Kr, $^{135}$Cs, $^{147}$Pm, $^{153}$Gd, $^{163}$Ho, $^{179}$Ta, which present sufficiently long half-lives to prepare a target ($\simeq$1~y) combined with the emission of no $\gamma$-rays or only low-energy $\gamma$-rays which could be efficiently shielded. The list of studied isotopes, their main properties, challenges for the TOF measurement and complementarity with activation are summarized in Table~\ref{tab:LimitsBranchings}. 

\begin{table}[h!] 
\caption{Summary of s-process branching points included in the feasibility study for TOF (n,$\gamma$) measurement in the upgraded n\_TOF EAR2~\cite{Lerendegui:25b}.\label{tab:LimitsBranchings}}
\begin{tabularx}{\textwidth}{CCCCC}
\toprule
\textbf{Isotope}	& \textbf{T$_{1/2}$}	& \textbf{A (\SI{1e17}{at})} & \textbf{Decay} & \textbf{Activation (T$_{1/2}$ product)}\\
\midrule
$^{81}$Kr   &  \SI{2.29E+05}{y}   & 9.60~kBq & \g:~275 keV, EC                     & No \\ 
$^{135}$Cs  &  \SI{2.30E+06}{y}   & 956~Bq   & No~\g, Q$_{\beta}$=286~keV         & Yes (13~d)\\ 
$^{147}$Pm  &  2.62 y             & 839 MBq  & No~\g, Q$_{\beta}$=224~keV         & Yes (5.4~d)\\ 
$^{153}$Gd  &  240 d              & 3.4 GBq  & \g:~<100 keV, EC                    & No\\ 
$^{155}$Eu  &  4.68 y             & 470 MBq  & \g:~86,105 keV, Q$_{\beta}$=252~keV & Yes (15.19 d) \\ 
$^{163}$Ho  &  \SI{4.57E+03}{y}   & 481 kBq  & No~\g, EC                           & Yes (30~min) \\ 
$^{179}$Ta  &  1.82~y             & 1.21 GBq & No~\g, EC                           & Yes (8~h)\\  
\bottomrule
\end{tabularx}
\end{table}
 
\begin{figure}[!h]
 \centering
  \includegraphics[width=0.495\columnwidth]{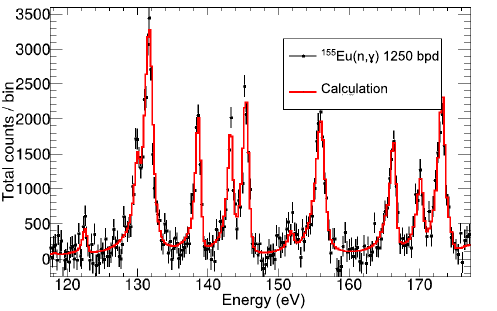}
  \includegraphics[width=0.495\columnwidth]{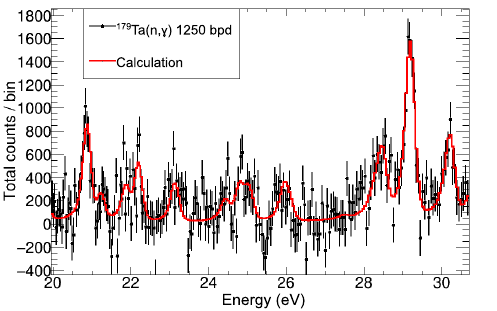}
 \caption{Expected capture yield with 1250 bins per decade (bpd) after the background subtraction compared to a the expected yield based on the simulated resonance parameters for a $^{155}$Eu sample of 5$\times10^{17}$ atoms (left) and a $^{179}$Ta sample of 10$^{18}$ atoms.}
 \label{fig:Expected_Results_155Eu_179Ta}
\end{figure}

 To study the feasibility of such experiments with the current sensitivity of EAR2, we have carried out simulations of the anticipated experimental outcomes for different sample masses ranging from 5$\times$10$^{16}$ atoms up to 5$\times$10$^{18}$ atoms. To realistically predict the statistical uncertainties we have implemented a Monte Carlo (MC) resampling method and assigned a realistic number of protons to the sample (corresponding to 30~days of measurement) and the background characterization (20~days) measurements. The isotopes of this study do not present high energy $\gamma$-ray activity and, thus, the major background source is expected to be the neutron-beam related background similarly to the situation of the recent $^{94}$Nb and $^{79}$Se measurements. Fig.~\ref{fig:Expected_Results_155Eu_179Ta} shows the expected capture yield for the potential  $^{155}$Eu(n,\g) and $^{179}$Ta(n,\g) with samples of, respectively \SI{5e17} and \SI{1e18} atoms. Here individual resonance parameters were taken from JEFF-3.3 and have been randomly generated based on average resonance parameters.

\begin{figure}[!h]
 \centering
  \includegraphics[width=0.8\columnwidth]{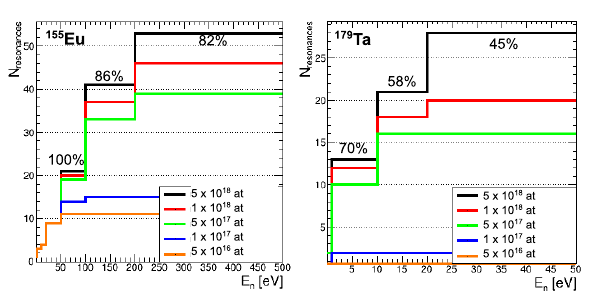}
 \caption{Expected number of cumulative observed s-wave resonances in different neutron energy ranges for $^{155}$Eu and $^{179}$Ta (4 $\sigma$ detection threshold). The cumulative fraction of observed resonances at the end of the energy range is provided (only for the thickest sample). Samples of 10~mm diameter and different masses are compared.}
 \label{fig:ObsRes_155Eu_179Ta}
\end{figure}

Based on the simulated experiments, we have quantified the expected number of observable resonances and the maximum neutron-energy range accessible for their identification. For each resonance, the probability of observation was evaluated using a statistical detection criterion following the methodology adopted in recent experiment preparation studies~\cite{Se79Proposal}. The results, shown in Fig.~\ref{fig:ObsRes_155Eu_179Ta} for the two representative isotopes, indicate that with the current signal-to-background ratio at n\_TOF-EAR2, samples containing \SI{1e17}--\SI{5e18} atoms (depending on the isotope), corresponding to sample masses of about 30--1500~$\mu$g, would be required to observe a sufficient number of resonances to allow a reliable determination of the average resonance parameters needed for the extrapolation of the MACS. In practice, at least 10--15 observed resonances are required for this purpose~\cite{Guerrero:2020,Casanovas:24,Balibrea:26}. Nonetheless, such sample masses are often beyond currently attainable quantities or cannot be produced with sufficient isotopic purity, highlighting the need for improved radiochemical production methods~\cite{Schumann:10}. 
The study also indicates that the signal-to-background ratio achievable in these TOF measurements would not allow a direct determination of the cross section in the unresolved resonance region (URR), which represents a large fraction of the neutron-energy range of astrophysical interest. For several of the studied isotopes --- $^{135}$Cs, $^{147}$Pm, $^{155}$Eu, $^{163}$Ho, and $^{179}$Ta --- a complementary activation measurement could therefore be envisaged (see Table~\ref{tab:LimitsBranchings}). Activation measurements will now be possible at CERN after the deployment of the n\_TOF-NEAR station and the proposal of the future n\_ACT activation facility discussed in the following.

As discussed along this section, direct capture measurements in the full energy range of stellar interest via TOF still require further optimization of the sensitivity. In particular, an improved SBR to access smaller sample mass ($\leq$100~$\mu$g), would be required. Starting from the state-of-the-art (n,\g) setup based on sTED detectors in a compact ring configuration~\cite{Lerendegui:23_NPA}, the n\_TOF collaboration has launched a series of optimization campaigns of the signal-to-background ratio (SBR) at EAR2~\cite{INTC_SBR_Opt,INTC_SBR_Opt2}. The efforts split in two directions: maximizing the detected (n,\g) events and reducing the beam-induced background.

\begin{figure}[!h]
 \centering
  \includegraphics[width=1.0\columnwidth]{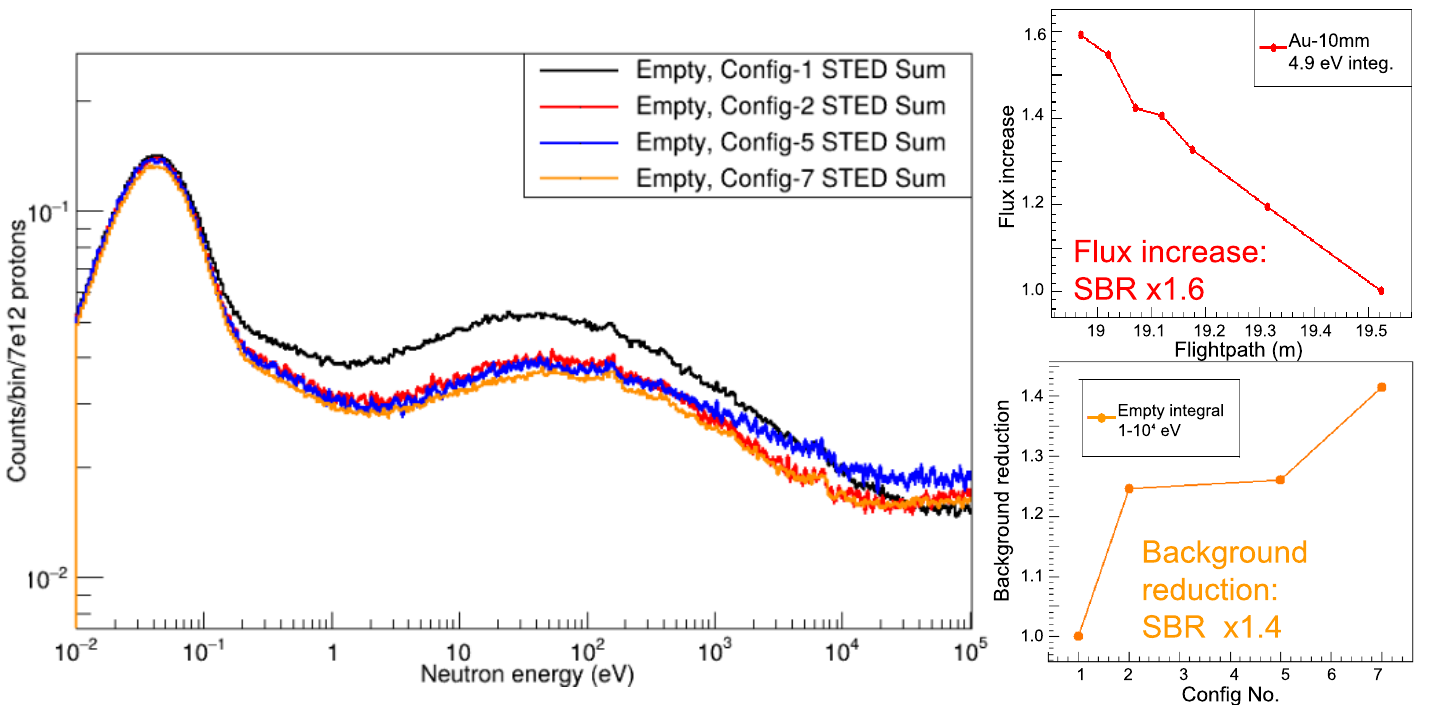}
 \caption{Left: \textit{Empty} background in n\_TOF EAR2 measured with the array of 9 sTED detector. The reference level (black) is compared with improved shielding configurations that provided an absolute reduction of the background. Left: Summary of the signal-to-background (SBR) gain obtained due to an increase in the flux intercepted by the sample (top) and due to the reduction of the empty sample background (bottom).}
 \label{fig:ImprovementSBR}
\end{figure}

The signal can be enhanced, by increasing the fraction of the neutron beam impinging on the sample, the so-called Beam Interception Factor (BIF). Recent results, shown in Fig.~\ref{fig:ImprovementSBR}, demonstrate that as a consequence of the spatial divergence of the beam in EAR2~\cite{Pavon:25}, the BIF can be enhanced by shortening the flight-path (L) in the experimental hall of EAR2. An improvement in the SBR by a factor of 1.6 was found for 10~mm diameter samples when bringing the sample and detectors 60~cm closer to the second collimator without any negative impact on the background. The (n,\g) sensitivity can be further enhanced by increasing the efficiency for detecting \g-rays emitted from the sample. This objective has motivated further R\&D efforts toward larger segmented arrays of \cd~detectors~\cite{Mendoza:23} or novel compact solid scintillators with higher intrinsic efficiency~\cite{Balibrea:25}. 

The remaining critical aspect for further improving the SBR is the reduction of the beam-induced background unrelated to the sample, so-called \textit{Empty} background in Fig.~\ref{fig:ImprovementSBR}. This background is intrinsic to the facility and depends largely on the effectiveness of the shielding against the uncollimated neutrons. The facility’s unique combination of a short flight path and presence of high energy neutrons up to 350~MeV complicates the design of a collimation system focused on the reduction of the background in (n,\g) experiments. The results of the recent optimization campaigns showed that, by improving the neutron shielding after the second collimator, the background could be reduced by a factor of 1.45 (see Fig.~\ref{fig:ImprovementSBR}). The combined increase in the (n,\g) signal and reduction of the background thus resulted in a 2.5-fold improvement in the SBR. These promising results have motivated a more detailed design study of a new shielding system aimed at achieving further substantial gains in sensitivity after LS3 (2026--2028). 

\subsection{The n\_TOF-NEAR: a high flux activation facility}

 As it was introduced in Sec.~\ref{Sec:Methods}, the activation technique shows an unsurpassed sensitivity and represents an advantage for the measurement of very small sample masses, where the limited signal-to-background ratio would represent an important limitation for the TOF measurement. In practice, combining neutron TOF and activation measurement, when feasible, may deliver complementary and more accurate information on a specific cross section, see e.g.~\cite{Guerrero:19,Guerrero:2020}. 
 
As already mentioned above, following these premises, the n\_TOF facility expanded in 2021 its experimental capabilities with the deployment of the NEAR station~\cite{Patronis:23}. Located at a distance of only 2.5~m from the lead spallation target, it features 2 orders of magnitude higher neutron flux than EAR2. The high flux makes it well suited for activation measurements on extremely small mass samples ($\leq$1$\mu$g) and on radioactive isotopes which, as discussed in Sec.~\ref{Sec:Advances_Limits}, are often not feasible via TOF. Conceptual MC simulations~\cite{Patronis:23,Domingo:25} showed the possibility of using moderating materials (e.g. Al-oxide, Be) followed by B$_4$C filters for producing quasi-Maxwellian neutron-energy spectra over a broad range of temperatures between about 1 and 100~keV, making this facility unique when compared with stellar neutron beams at $kT$~=~25~keV. After a few years of development and commissioning ~\cite{StamatiND:22}, the energy-tuning setup is now fully implemented after the recent installation of a 20~cm thick Al-oxide moderator~\cite{Aberle_INTC_2026} which smoothens the flux and suppresses the undesired MeV-neutrons component~\cite{Domingo:25}. The activated samples are then measured at the Gamma-ray spectroscopy Experimental ARea (GEAR) of n\_TOF, which is at the moment based on a CANBERRA HPGe detector GR5522 (63\% efficiency) supplemented with a low-background shielding~\cite{Patronis:23}. 

Besides the possibility to exploit the complementarity between TOF and activation, illustrated with the recent $^{146}$Nd measurement (see Sec.~\ref{Sec:Highlights_nTOF}), NEAR aims at pushing the limits for the direct capture measurements on radioactive nuclei not accessible by TOF. The first unstable isotope of astrophysical interest proposed for an activation MACS measurement at NEAR is $^{135}$Cs (T$_{1/2}$~=~ 2 Myr)~\cite{INTC_Cs135_2}. The stellar neutron-capture rate of $^{135}$Cs is of particular relevance as a part of the temperature-dependent s-process branching at A = 134-135~\cite{Taioli:22}, which fixes the branching ratio between the two s-only $^{136}$Ba and $^{134}$Ba, well characterized from SiC in presolar grains~\cite{Palmerini:21}. The neutron capture of $^{135}$Cs at $kT$~=~25~keV was already measured at FZK~\cite{Patronis:04} and therefore this measurement could be a good benchmark case for the performance of the new facility. In addition, at NEAR the MACS could also be completed for lower neutron energy ranges around $kT$~=~8~keV, where presently there is no experimental information available.

\begin{figure}[!h]
 \centering
  \includegraphics[width=0.95\columnwidth]{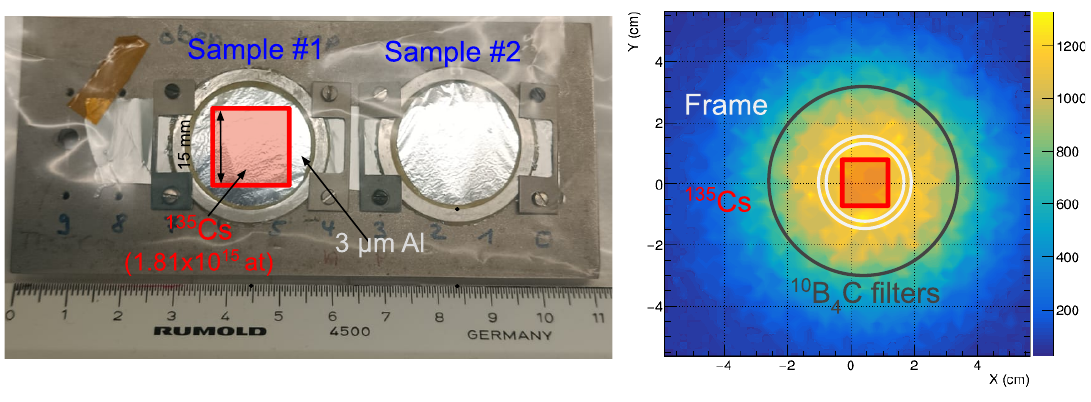}
 \caption{Left: Sample backing where $^{135}$Cs was implanted on the holder of the ISOLDE SSP chamber. Right: Sketch of the sample inserted in the filters overlaid with the neutron beam profile at NEAR. The color scale represents the neutron flux in a.u.}
 \label{fig:Cs135}
\end{figure}

A suitable sample of $^{135}$Cs of \SI{1.81e15} atoms, sketched in Fig.~\ref{fig:Cs135}, has been ion-implanted on a 3~$\mu$m thick aluminum backing in the Solid State Physics (SSP) chamber of ISOLDE~\cite{INTC_Cs135}. The $^{135}$Cs beam was produced and mass-separated using the General Purpose Separator using as an ion source a commercial $^{137}$Cs liquid source of 9~MBq featuring an isotopic ratio $^{135}$Cs/$^{137}$Cs=0.72. The sample backings were designed to fit in the B$_4$C discs used for the aforementioned energy filtering at NEAR, where the beam dimensions are much larger than the 15$\times$15mm$^2$ implanted area (see Fig.~\ref{fig:Cs135}). After the irradiation, the decay of the activation product,$^{136}$Cs (T$_{1/2}$~=~13 d), could be measured at the GEAR station or in other high-efficiency HPGe-based setups (available at CERN). This experiment would show the potential synergy to perform activation measurements at NEAR on small samples of highly isotopically enriched samples produced in the nearby ISOLDE facility.

\section{New ideas and future projects}\label{Sec:Future}

While the recent upgrades at n\_TOF have substantially improved present capabilities, fully addressing the remaining challenges requires the development of novel experimental concepts and facilities. This section outlines complementary experimental concepts that aim to overcome the limitations identified in Sec.~\ref{Sec:Advances_Limits} by either increasing neutron flux, extending energy coverage, or developing new methods to access short-lived nuclei.

\subsection{New high-flux activation facilities at CERN}
\subsubsection{The n\_TOF-NEAR facility with CYCLING}

To date, activation measurements at NEAR consist of week-long irradiations followed by manual retrieval and transport to the GEAR. In this single-activation scheme, the time required to cool-down the NEAR bunker after the end of the irradiation (6~h) and transport the sample to the offline detector, imposes an intrinsic limit on the minimum half-life of the (n,$\gamma$) product. This limitation could be overcome with the installation of a CYCLIc activation station for (N,G) experiments ({\it \cycling}) at \near{}~\cite{loi_background,Lerendegui:25b}, the repetition of a short irradiation, rapid transfer to a detector located nearby, measurement of the decay and transport back to the irradiation position~\cite{Beer_Cyclic}. Such a remote sample transport system at \near{} would also allow the remote retrieval of any irradiated samples, thereby reducing the duration and frequency of beam stops to access \near.

Depending on the duration of the transport of the sample to the detector, half-lives of the order of seconds or minutes would become accessible. There are several isotopes of great interest for stellar nucleosynthesis studies that could be investigated with the proposed CYCLING station. Some of them are stable nuclei, such as $^{19}$F, which leads to the formation of $^{20}$F(T$_{1/2}$~=~11~s)~\cite{loi_background}. Among the s-process branching nuclei listed in Table~\ref{tab:LimitsBranchings}, the activation measurements of $^{163}$Ho($n,\gamma$)$^{164}$Ho(T$_{1/2}$~=~30~min) and $^{179}$Ta($n,\gamma$)$^{180}$Ta(T$_{1/2}$~=~8~h) would also benefit of a cyclic scheme at NEAR. In addition to s-process cases, ($n,\gamma$) measurements on isotopes of relevance for the intermediate ($i$-) neutron capture process~\cite{Cowan77}, such as $^{137}$Cs($n,\gamma$)$^{138}$Cs(T$_{1/2}$~=~33 min) or $^{144}$Ce($n,\gamma$)$^{145}$Ce(T$_{1/2}$~=~2.8 min)~\cite{Domingo:22,Lerendegui:25b}, could be accessible for the first time with CYCLING.

\begin{figure}[!h]
 \centering
  \includegraphics[width=0.7\columnwidth]{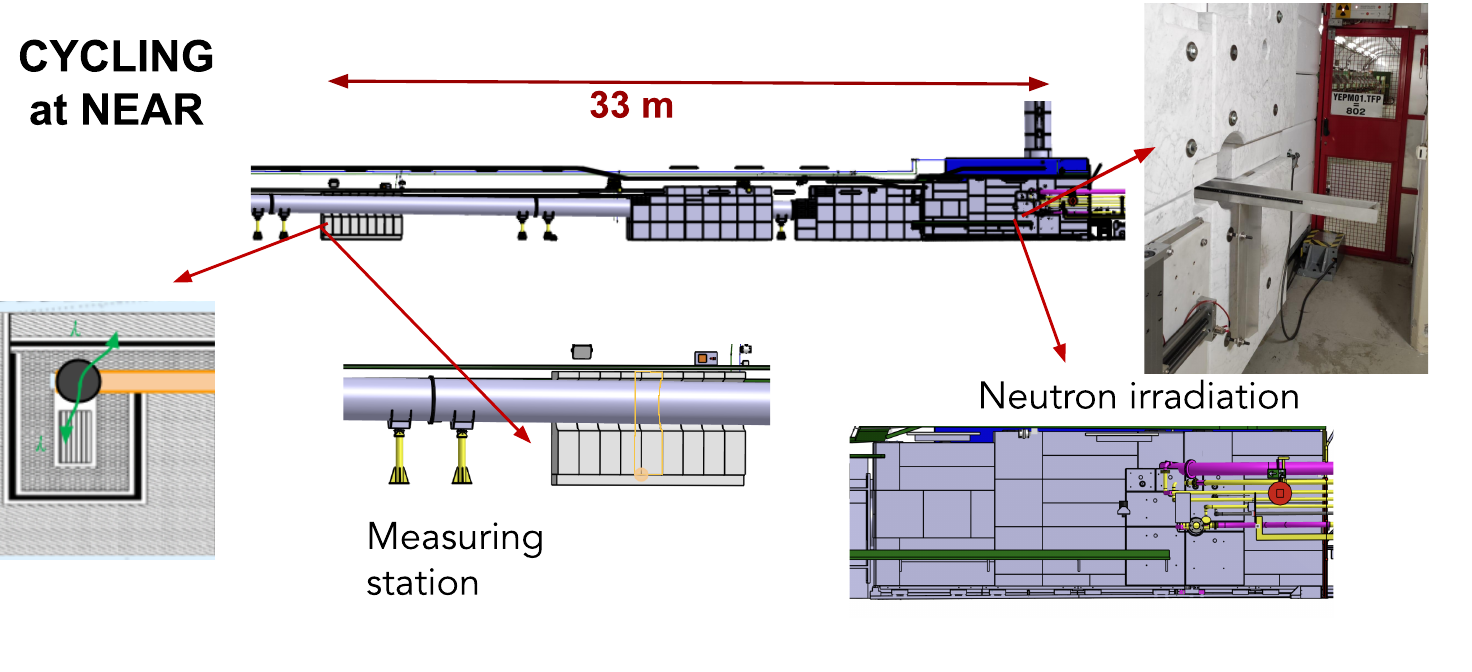}
 \caption{Sketch of the CYCLING project showing the irradiation point for activations at NEAR (a-NEAR) and the measuring station, proposed to be at 33~m in the tunnel towards EAR1, that would be connected by a remote transport system.}
 \label{fig:Cycling_sketch}
\end{figure}

The feasibility of this station depends mainly on the capability to operate active detectors in the harsh radiation environment near the \near{} bunker and thus the design of the measuring station requires a good knowledge of the expected neutron and $\gamma$-ray fields. With this goal, several campaigns have been carried out to assess the mixed neutron-gamma radiation field at several positions in the \near{} and validate the first simulation results~\cite{Lerendegui:25b}. The sensitivity to measure the decay of activated samples has been explored using a portable decay station consisting on a cylindrical 1.5”×1.5” LaBr$_3$(Ce) Canberra detector with shielding for neutrons (borated polyethylene) and $\gamma$-rays (steel) that has been placed in various positions around the NEAR facility. Based on these campaigns, the preferred location, which provides a sufficiently low background, is 33~meters away from the irradiation point in the downstream tunnel towards EAR1 and shielded from the neutrons a large concrete shielding, see Fig.~\ref{fig:Cycling_sketch}.

\begin{figure}[!h]
 \centering
  \includegraphics[width=0.51\columnwidth]{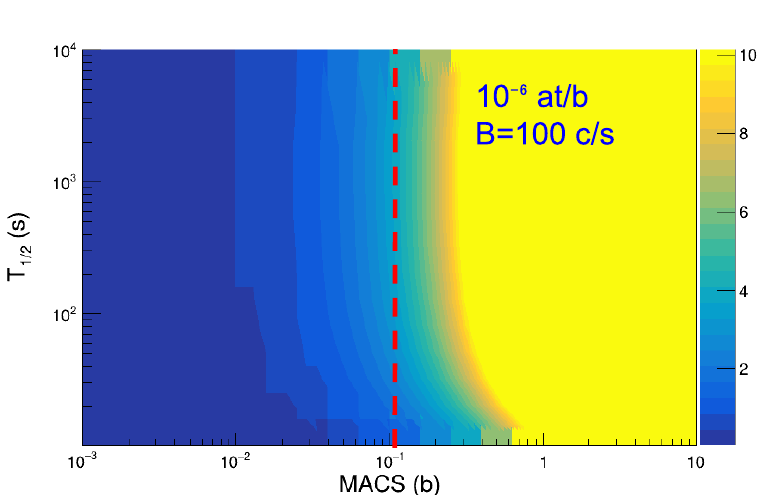}
  \includegraphics[width=0.48\columnwidth]{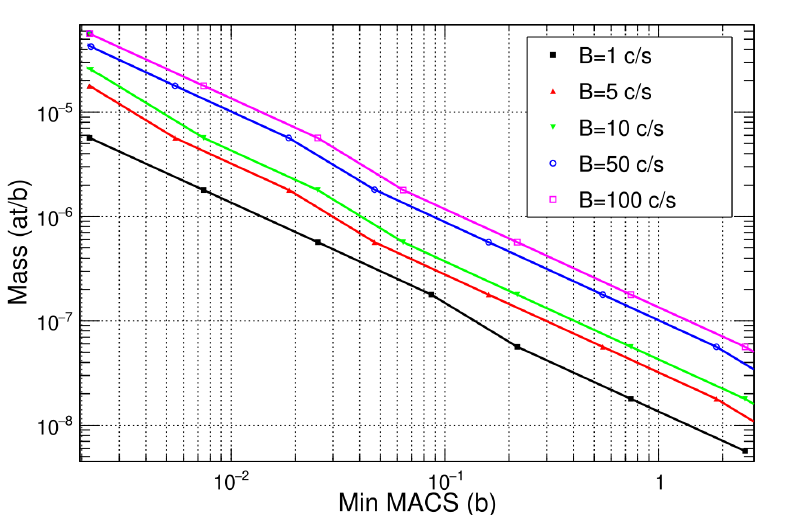}
 \caption{Left: Simulated statistical significance (colour scale in $\sigma$) for the detection of decay a $\gamma$-ray in a cyclic activation scheme as a function of the MACS$_{30}$ and the half-life of the (n,$\gamma$) product nucleus. These results correspond to the scenario for the background of $\simeq$100~c/s in a HPGe. The red dashed line indicates the minimum MACS measurable (at the significance level of 4$\sigma$). Right: Mass required as a function of the minimum detectable MACS$_{30}$ (Min MACS) for different background levels in the final measuring station with an optimized shielding.}
 \label{fig:Cycling_sensitivity}
\end{figure}

Based on the signal-to-background ratio achieved in the experimental campaign, a systematic study of cases that would be measurable in CYCLING has been carried out. A numerical simulation of a cyclic activation scheme was performed assuming a fixed transport time of 15~s, irradiation and counting times of 1.5$\times T_{1/2}$ and a time-averaged neutron flux of \SI{3.07e7}{n/s}, corresponding to the energy-filtered flux integrated over a sample of 10~mm radius. The feasibility has been studied as a function of the (n,$\gamma$) product $T_{1/2}$, the MACS$_{30}$ and the areal density of the activated sample. From the results, we have evaluated the statistical significance for observation of a photo-peak corresponding to the $\gamma$-ray decay of the product nucleus. Fig.~\ref{fig:Cycling_sensitivity} shows the results for this baseline scenario -- based on the background level found in the exploratory campaigns of $\approx$100 c/s below the 412~keV photo-peak of the $^{198}$Au decay. The results show that the $T_{1/2}$ of the (n,$\gamma$) product does not impact the feasibility, except for cases where it is comparable to the transport time. The study concluded that the assumed background level would allow one to measure MACS$_{30}$ in the range of 10--1000~mb with samples of 10$^{-7}$~-~10$^{-5}$~at/b, corresponding to 10$^{17}$~-~10$^{19}$~atoms (see right panel of Fig.~\ref{fig:Cycling_sensitivity}). The final attainable sensitivity would depend on the level of background after an optimized design of the built measuring station.

For unstable isotopes, the background will in most cases be dominated by the intrinsic activity of the sample itself, and the feasibility of such experiments will depend strongly on the decay-energy spectrum and the detector counting-rate capabilities, thus requiring a case-by-case optimization of the setup. Among the most challenging cases discussed above is $^{137}$Cs: a sample of $\sim$\SI{1e17}{atoms} would correspond to an activity of about 70~MBq dominated by the 662~keV $\gamma$-ray line. In this scenario, a viable strategy could rely on high count-rate detectors such as LaBr$_3$(Ce), combined with a relatively high energy threshold or the use of a lead absorber to suppress the intense 662~keV radiation, thereby enabling the detection of the higher-energy ($\geq$1~MeV) $\gamma$ rays from the decay of $^{138}$Cs.

The implementation of CYCLING would require the installation of a remote transport system that is able to cover the 30~m distance in tens of seconds. The design phase of the project has been launched and, if a viable solution is found, it could be developed and installed at CERN before the end of LS3.

\subsubsection{The n\_ACT at BDF facility}

To expand further on the large potential of the activation technique and on its complementarity with TOF experiments, the n\_ACT facility at the SPS Beam Dump Facility (BDF) has been proposed at CERN~\cite{BDF_Proposal}. n\_ACT will be a high-flux neutron activation station integrated into the BDF target, which can be operated parasitically to the Search for Hidden Particles (SHiP) experiment. The high intensity neutron fields produced by spallation reactions of the 400 GeV/c, \SI{4e13}{p/pulse} proton beam from the SPS accelerator in the tungsten target, can be exploited for accurate neutron activation measurements.

\begin{figure}[!h]
 \centering
  \includegraphics[width=1.0\columnwidth]{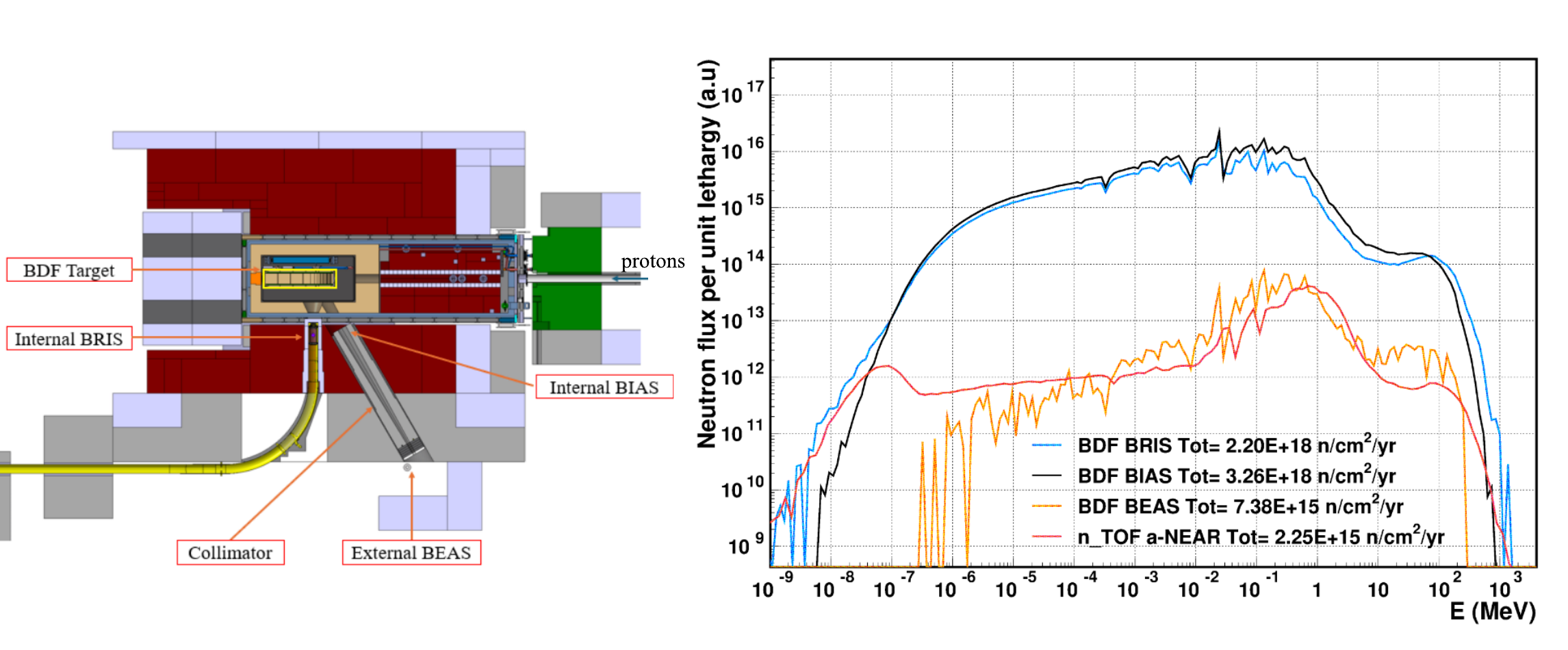}
 \caption{Left: Layout of the proposed n\_ACT Activation Stations and the main components of the BDF target station. Right: Neutron spectra at BDF irradiation stations (BRIS, BIAS and BEAS) compared with n\_TOF a-NEAR. Figure taken from Ref.~\cite{BDF_Proposal}. }
 \label{fig:BDF}
\end{figure}

n\_ACT will comprise three complementary activation stations integrated into the BDF target complex (see Fig.~\ref{fig:BDF}) offering ultra-high-flux close-target neutron fields for long-term irradiations (BIAS), a rabbit system to connect with an external radiochemistry and spectroscopy lab (BRIS), and a collimated external neutron beam (BEAS), thereby covering the full range of activation needs. As shown in Fig.~\ref{fig:BDF}, BIAS and BRIS would provide neutron fluxes up to three orders of magnitude higher than that of a-NEAR. n\_ACT would profit from the filtering method developed at NEAR to tailor the neutron spectra to different stellar $kT$ values (2--60 keV). Being at CERN, this facility would profit from the unique complementarity with n\_TOF --- energy-resolved via TOF combined with high-flux activation --- and proximity to ISOLDE for producing short-lived isotopes and should enable the world-first measurements on several rare and unstable nuclei. 

\subsection{TOF-DONES: A time-of-flight facility at IFMIF-DONES}

\begin{figure}[!h]
 \centering
  \includegraphics[width=0.7\columnwidth]{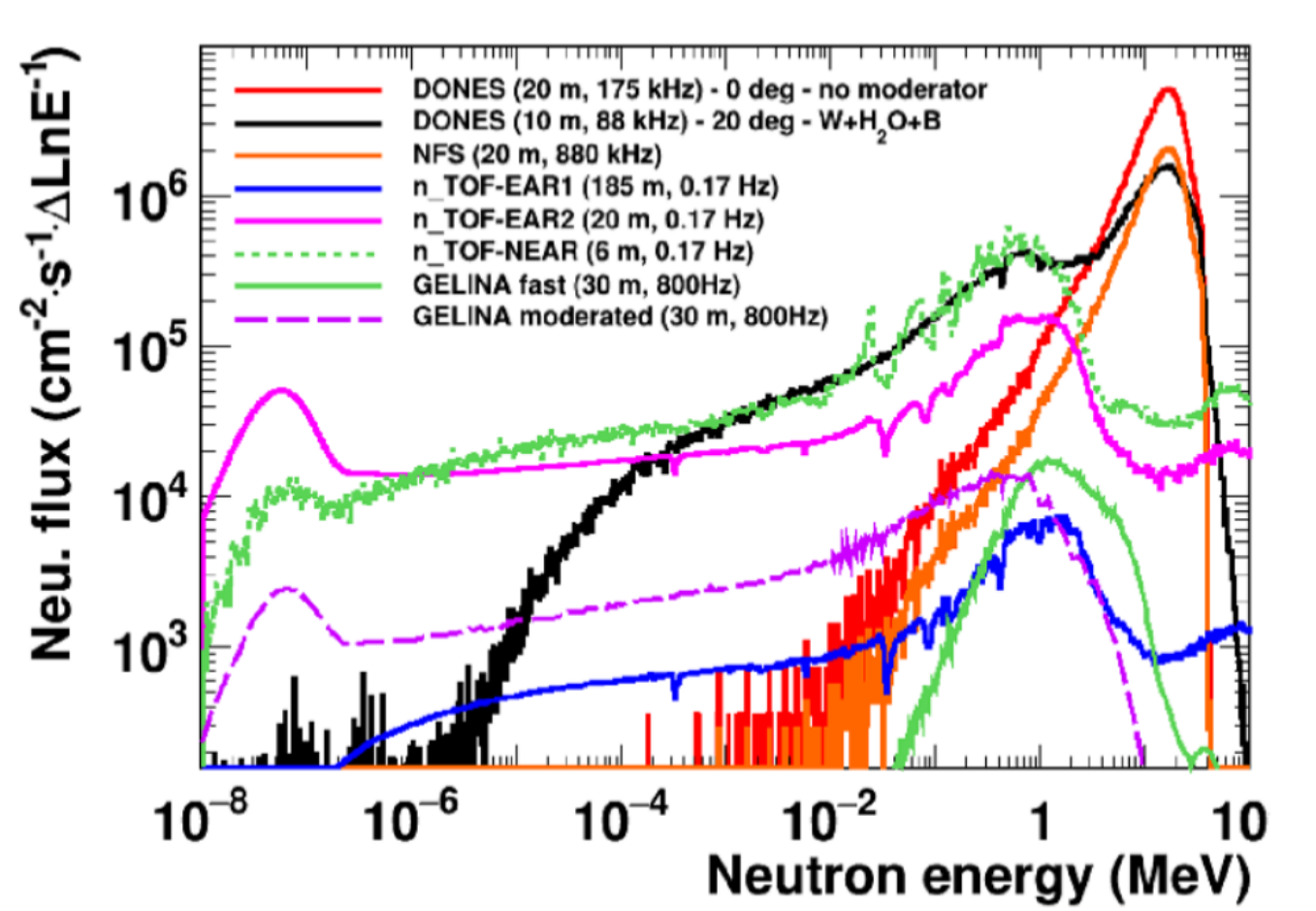}
 \caption{Average neutron flux at different TOF beamlines in TOF-DONES compared to other existing TOF facilities. Figure taken from Ref.~\cite{DONES}.}
 \label{fig:TOFDOnes}
\end{figure}

As discussed in Sec.~\ref{Sec:Advances_Limits}, pushing the limits of TOF capture measurements requires developing new facilities that can substantially improve the performance of the existing ones. With this aim, a TOF facility at the IFMIF-DONES facility, is being designed~\cite{DONES}. At TOF-DONES, the extraction of 0.1\% of the 40-MeV deuteron beam bunched in 5.3 ns pulses with a maximum repetition rate of 175~kHz, would be used for driving one of the most intense neutron TOF facilities. The performance in terms of time-averaged neutron flux of two TOF-DONES beam lines when compared with existing leading neutron TOF facilities is shown in figure~\ref{fig:TOFDOnes}. 

The TOF-DONES facility is projected to have a graphite neutron production target and several flight paths with different moderation, that would provide similar or even higher neutron flux of the currently most intense TOF facilities above 10~keV (see Fig.~\ref{fig:TOFDOnes}). The unmoderated neutron spectrum would enable to have a world-leading intensity of fast neutrons, higher than that of the NFS facility at GANIL. In addition, one of the foreseen beamlines would have a moderated spectrum that would extend the neutron energies down to the astrophysical range of interest with an average flux larger than that of n\_TOF EAR2.

\subsection{Inverse kinematics with storage rings and neutron targets}

Direct capture measurements using neutron beams and targets of the isotope of interest will never allow accessing short-lived nuclei (T$_{1/2}~\leq$~days) of relevance for the s-, i- and r-process. To tackle these cases, a promising approach for the long-term future consists of performing direct neutron-capture reactions in inverse kinematics by circulating radioactive ions in a storage ring intersecting a localized free-neutron target \cite{Reifarth:14,Reifarth:17}. In this concept, ions repeatedly traverse a confined neutron field, dramatically enhancing the effective luminosity and enabling direct measurements on short-lived nuclei that cannot be prepared as conventional samples. With the inverse kinematics method, direct measurements of neutron-capture cross sections at different stellar temperatures could be performed by changing the ion-beam energy.



The first conceptual realization of this idea was based on the use of a nuclear reactor~\cite{Reifarth:14}. A more feasible approach was proposed later based on a spallation neutron source \cite{Reifarth:17}. In the latter high-energy protons impinge on a tungsten target surrounded by a heavy-water moderator, producing a high intensity, moderated neutron field overlapping the ion beam pipe of a storage ring. In parallel, a new initiative focuses on developing a compact cyclotron-driven neutron target, where neutrons are produced via low-energy $^{9}$Be(p,xn) reactions and subsequently moderated and confined around the storage-ring beam pipe using optimized moderator and reflector assembly~\cite{Tarifeno:26}. This design enables a fully integrated and scalable neutron target system aimed at compatibility with existing or future storage rings. 
Building on this concept, several facilities are actively developing projects in this direction. At TRIUMF-ISAC, the proposed TRISR storage ring aims at integrating a compact neutron source directly within the ring lattice, enabling circulating radioactive ions from the ISOL facility to interact with moderated neutrons~\cite{Dillmann:23}. Another neutron target project aims to validate the concept at CRYRING@ESR (GSI) (see Fig.~10 of Ref.~\cite{Tarifeno:26} for the proposed implementation). The overall performance would then be further boosted by several orders of magnitude~\cite{Tarifeno:26} by integrating this neutron target in specially designed ion storage such as TRISR at TRIUMF~\cite{Dillmann:23} or ISR at ISOLDE~\cite{Grieser:12}.

Examples for neutron capture measurements of astrophysical interest that could become accessible for a direct measurement with this concept are some of the s-process branching points which remain unmeasured, e.g. $^{147}$Nd~(T$_{1/2}$~=~11~d), $^{148}$Pm~(T$_{1/2}$~=~5.4~d), $^{160}$Tb~(T$_{1/2}$~=~72~d) and $^{170}$Tm~(T$_{1/2}$~=~128~d) as well as shorter-lived isotopes with high impact in the i-process abundances, such as $^{66}$Ni~(T$_{1/2}$~=~55~h), $^{72}$Zn~(T$_{1/2}$~=~47~h), $^{135}$I~(T$_{1/2}$~=~6.6~h),
$^{141}$Ba~(T$_{1/2}$~=~18 min),$^{141}$La~(T$_{1/2}$~=~3.9~h) and $^{153}$Sm~(T$_{1/2}$~=46~h).

\section{Summary and outlook}
Neutron-capture reactions remain one of the fundamental nuclear-physics inputs governing the synthesis of heavy elements in stars. Despite more than five decades of experimental effort, significant challenges persist in achieving the level of accuracy and completeness now demanded by modern astrophysical observations. High-precision isotopic measurements from presolar SiC grains, improved spectroscopic stellar abundances, and increasingly sophisticated stellar models require neutron-capture cross sections with uncertainties at or below the 5\% level over the full stellar energy range. In addition, key branching-point isotopes and nuclei relevant for the intermediate (i-) process remain largely unexplored experimentally.

In this work, we have reviewed recent advances in direct neutron-capture measurements at CERN n\_TOF, highlighting both achievements and present limitations. Time-of-flight (TOF) experiments at n\_TOF have significantly improved the knowledge of cross sections for s-only isotopes and bottleneck nuclei, such as $^{154}$Gd, $^{140}$Ce and $^{209}$Bi, directly impacting stellar abundance predictions. In parallel, substantial progress has been achieved in the measurement of unstable branching-point nuclei, including first-time capture measurements of isotopes such as $^{94}$Nb and $^{79}$Se. These results have been driven by the high luminosity of EAR2, the development of advanced detection systems such as sTED and i-TED, and continuous optimization of the signal-to-background ratio.

At the same time, systematic feasibility studies show that even with state-of-the-art TOF facilities like n\_TOF EAR2, measurements of s-process branching-point isotopes remain limited by sample availability, background conditions, and restricted energy coverage. In several cases, only the resolved resonance region can be accessed with realistic sample masses, leaving the Maxwellian averaged cross sections at higher stellar temperatures insufficiently constrained. In this context, the complementarity between TOF and activation techniques emerges as a central strategy for the next generation of neutron-capture experiments. The commissioning of the high-flux n\_TOF-NEAR activation station opens new possibilities for measuring extremely small or radioactive samples and for tailoring quasi-Maxwellian neutron spectra over a wide range of stellar temperatures. The case of $^{146}$Nd illustrates this synergistic approach: a high-resolution TOF measurement in the resonance region, combined with activation data at different $kT$ values, enables a consistent re-evaluation of the stellar rate and directly addresses discrepancies between s-process models and presolar grain data.

Looking ahead, further sensitivity gains are expected from future optimization of n\_TOF setups during CERN LS3. The proposed CYCLING station at n\_TOF NEAR would extend activation measurements to short-lived reaction products. In parallel, the n\_ACT project at the CERN Beam Dump Facility and the development of TOF-DONES would provide substantially higher neutron fluxes and complementary energy coverage opening access to new s- and i-process cases. In the even longer term, inverse-kinematics approaches using storage rings in which unstable ion beams circulate through localized neutron targets offer a novel path toward direct measurements on short-lived nuclei (T$_{1/2}~\leq$~days) that are inaccessible today. Continued progress along these lines will be essential for reducing the remaining nuclear-physics uncertainties and for fully exploiting the precision of modern astrophysical observations in unraveling the origin of heavy elements.

\vspace{6pt} 

\authorcontributions{Conceptualization, J.L.M ; methodology, J.L.M; formal analysis, J.L.M ; investigation, n\_TOF-Collaboration; resources, n\_TOF-Collaboration ; data curation, J.L.M.; writing---original draft preparation, J.L.M.; writing---review and editing, C.D.P; Supervision: C.D.P; project administration: C.D.P, J.L.M; funding acquisition: C.D.P, J.L.M,. All authors have read and agreed to the published version of the manuscript.}

\funding{This research received no external funding apart from those included in the acknowledgments section.}

\dataavailability{The datasets generated during and/or analyzed during the current study are available from the corresponding author on reasonable request.} 

\acknowledgments{Part of this work was supported by the European Research Council (ERC) under the European Union’s Horizon 2020 research and innovation programme ERC-COG No. 681740 and ERC-STG No. 677497, European H2020-847552 (SANDA) and EURO-LABS (grant agreement No. 101057511). The authors acknowledge support of the Spanish MCIN/AEI 10.13039/501100011033 under grants Severo Ochoa CEX2023-001292-S, PID2022-138297NB-C21, PID2019-104714GB-C21. J.L.M acknowledges support of grant FJC2020-044688-I funded by MCIN/AEI/ 10.13039/501100011033 and European Union NextGenerationEU/PRTR, and grant CIAPOS/2022/020 funded by the Generalitat Valencia and the European Social Fund. Authors also acknowledge the funding agencies of the participating n\_TOF institutions and Infrastructure Access Agreement N° 35543/1/2019-1-RD-EUFRAT-GELINA.
}

\conflictsofinterest{The authors declare no conflicts of interest.} 





\begin{adjustwidth}{-\extralength}{0cm}

\reftitle{References}

\PublishersNote{}
\end{adjustwidth}
\end{document}